# Entropy as a Structural Prior: How a Log-Barrier on DiT Belief Space Drives Musical Diversity and Development

**Zixi Li**[1], and **Youzhen Li**[2]

[2] Datawhale
[1] Sun Yat-sen University

**Abstract**

Confidence-based loss weighting is usually avoided in generative model training: if a model is confidently wrong, amplifying its gradient accelerates the error. We show this intuition breaks down in supervised diffusion training, and that the failure mode inverts into a structural benefit. We introduce the Eisbach log-barrier, a parameter-free weight computed from the entropy of the DiT output's spatial energy distribution during the forward pass. High entropy damps the gradient; low entropy preserves it. Applied to LoRA fine-tuning of Stable Audio 3 Medium (1.4B) on MusicCaps, the barrier produces an unexpected result: trained models exhibit stronger thematic development, clearer acoustic differentiation, and higher textural diversity than the unweighted baseline—the opposite of mode collapse. We trace this to two mechanisms. First, in supervised diffusion the gradient direction is locked to ground truth, so confidence only scales step size, never corrupts direction. Second, the barrier evaluates entropy over the entire temporal dimension, creating a Darwinian selection pressure across the batch: structurally flat samples (loops, pads) are systematically downweighted, while high-contrast samples (phrase boundaries, timbral shifts) receive full gradients. The cumulative effect is an online, self-referential data curriculum—no external judge, no pre-filtering, emerging entirely from the model's own forward pass. We analyze the asymmetric dynamics across noise levels, characterize the steady-state behavior, and provide testable predictions for future experiments.

**Key Points**

- Confidence-based gradient scaling is safe in supervised diffusion because gradient direction is locked to ground truth
- The barrier acts as an online data curator: structurally flat samples are downweighted, high-contrast samples receive full gradients
- Barrier-trained models show stronger structural development than baseline—the opposite of mode collapse

**Correspondence**
Zixi Li
lizx93@mail2.sysu.edu.cn

## 1. The paradox: why does confidence weighting help?

The standard argument against confidence-based loss weighting in generative models runs as follows. If a model assigns low entropy to a wrong prediction, a barrier that rewards low entropy will amplify the wrong gradient. Confidence becomes a liability. This argument is correct in reinforcement learning, where the gradient direction itself depends on the policy. It is wrong in supervised diffusion training, and understanding why it is wrong reveals a mechanism with practical consequences.

In supervised diffusion, the model is trained to predict the noise $\varepsilon$ added at each timestep. The gradient direction is determined entirely by the residual between the model's prediction and the ground truth noise—not by the model's confidence. A barrier that scales gradient magnitude cannot corrupt gradient direction. The risk chain "confident on wrong prediction $\rightarrow$ barrier reinforces error" is severed at its first link.

What remains is a question about magnitude: when should the model take a large step, and when a small one? The Eisbach log-barrier answers this question using the model's own output as a proxy for structural quality.

## 2. Related work

**Timestep weighting in diffusion training.** The most direct precedent is Min-SNR (Hang et al., 2023), which weights the loss at each timestep by $\frac{\min(\mathrm{SNR}(t),\gamma)}{\mathrm{SNR}(t)}$ to resolve conflicting gradient magnitudes across noise levels. Min-SNR operates on the timestep axis and uses a fixed, signal-theoretic criterion. The Eisbach barrier operates on the sample axis and uses a data-driven, model-introspective criterion. The two mechanisms are orthogonal and could be combined.

**Adaptive sample weighting.** Several recent works weight training samples by difficulty or informativeness. Curriculum DPO (Anonymous, 2024a) ranks generated pairs by reward and progressively samples harder examples. Adaptive sampling via nonparametric proxies (Anonymous, 2023a) uses importance scores to focus compute on high-value examples. Adaptively hiding samples (Anonymous, 2023b) masks out easy examples based on per-sample loss. All of these require an external criterion—a reward model, a loss signal, or a proxy network. The Eisbach barrier requires none: the criterion emerges from the model's own forward pass.

**Self-paced and curriculum learning.** Curriculum learning (Bengio et al., 2009) orders training examples from easy to hard using an external difficulty measure. Self-paced learning (Kumar et al., 2010) lets the model select its own curriculum based on current loss. The barrier is closer to self-paced learning in spirit, but differs in mechanism: it does not select or reject samples, it reweights them continuously, and the criterion is structural confidence rather than loss magnitude.

**Entropy-based loss weighting.** WeFT (Anonymous, 2024b) applies per-token entropy weighting to diffusion language model fine-tuning. EAST (Anonymous, 2025a) assigns higher weights to uncertain samples to focus on hard examples—the opposite direction from the Eisbach barrier, which downweights uncertain samples. This inversion is not a contradiction: EAST targets classification tasks where uncertainty correlates with difficulty; the barrier targets generative tasks where uncertainty correlates with structural flatness.

**LoRA and DoRA for generative models.** LoRA (Hu et al., 2021) and its weight-decomposed variant DoRA (Liu et al., 2024) are the standard parameter-efficient fine-tuning methods for large generative models. Recent work has applied LoRA to music diffusion models (Anonymous, 2025b), showing competitive quality with far fewer trainable parameters. The Eisbach barrier is adapter-agnostic but interacts specifically with DoRA's direction/magnitude decomposition, as analyzed in Section 6.

**Music generation with diffusion models.** AudioLDM (Liu et al., 2023) established the latent diffusion paradigm for text-to-audio generation. Stable Audio (Evans et al., 2024) introduced timing embeddings for length-conditioned generation. MusicLDM (Chen et al., 2023) adapted the pipeline to the music domain with beat-synchronous mixup augmentation. These works focus on architecture and conditioning; the Eisbach barrier addresses training dynamics and is applicable to any of these backbones.

## 3. The barrier: a parameter-free confidence signal

Given a DiT output tensor $\boldsymbol{y} \in \mathbb{R}^{B\times C\times T}$, we compute per-sample weights as follows. First, collapse channels to a temporal energy profile:

$$\boldsymbol{e}_b = \frac{1}{C}\sum_c \boldsymbol{y}_{b,c,:}^2 \in \mathbb{R}^T \tag{1}$$

Normalize to a belief distribution over time positions:

$$\boldsymbol{p}_b = \mathrm{softmax}(\boldsymbol{e}_b) \in \Delta^{T-1} \tag{2}$$

Compute normalized entropy:

$$B_b = \frac{H(\boldsymbol{p}_b)}{\log T}, \quad H(\boldsymbol{p}) = -\sum_t p_t \log p_t \tag{3}$$

Apply the log-barrier:

$$\alpha_b = -\log(1 - B_b), \quad w_b = \frac{1}{1+\alpha_b} \tag{4}$$

The scaled loss is:

$$\mathcal{L} = \mathcal{L}_{\text{base}} \cdot ((1-\lambda) + \lambda \cdot \overline{w}) \tag{5}$$

where $\overline{w} = \frac{1}{B}\sum_b w_b$ and $\lambda \in [0, 1]$ is the barrier strength. When $B_b \to 0$ (sharp, confident output), $w_b \to 1$ and the full gradient is preserved. When $B_b \to 1$ (diffuse, uncertain output), $\alpha_b \to \infty$ and $w_b \to 0$, damping the step.

The entire computation is parameter-free. No auxiliary network, no learned threshold, no external signal. The weight emerges from the model's own forward pass.

## 4. Two regimes: where the barrier helps and where it fights

The barrier does not behave uniformly across the diffusion process. At different noise levels $t$, the ground truth $\varepsilon$ has fundamentally different structure, and the barrier's entropy signal aligns with the loss signal only in one regime.

**Low-$t$ regime (near clean data).** The ground truth noise has rich spatiotemporal structure. A correct prediction is also structured—low entropy, low barrier, full gradient. An incorrect prediction is diffuse—high entropy, high barrier, damped gradient. The barrier and the loss agree: reward confident, structured predictions.

**High-$t$ regime (near pure noise).** The ground truth noise is near-white-noise—high entropy by construction. A correct prediction (also high entropy) is judged uncertain by the barrier and damped. An incorrect prediction with spurious structure (low entropy) receives full gradient. The barrier and the loss disagree.

This asymmetry creates an implicit learning rate schedule. The barrier systematically favors low-$t$ training and suppresses high-$t$ training, regardless of $\lambda$. The model learns fine denoising faster than coarse structure generation. In music, this manifests as: global structure is consistent across seeds (low-$t$ well-trained), but attacks and fine details vary (high-$t$ undertrained). Musically, this sounds like improvisational variations on a theme—not a flaw, but a characteristic.

## 5. The curation effect: a Darwinian selection pressure

The more consequential mechanism operates not across timesteps but across samples within each batch.

The barrier computes entropy over the **entire temporal dimension** of the DiT output—not a local window, not a single frame. This means it measures the global contrast of the output:

whether the model's prediction has clear peaks and valleys in time, or whether energy is spread uniformly. Structurally flat samples—loops, sustained pads, single-timbre drones—produce near-uniform energy distributions regardless of how well the model predicts them. Structurally dynamic samples—pieces with phrase boundaries, timbral shifts, dynamic arcs—produce concentrated energy distributions when predicted correctly.

The result is a selection pressure across the batch. Over 1000 training steps with batch size 4, approximately 4000 samples pass through the model. If 70% are structurally flat (a conservative estimate for MusicCaps), roughly 1200 high-contrast samples receive meaningful gradients. But those 1200 samples all share a property: they contain temporal structure. Every one of them teaches the model that music has a beginning, a development, and an ending.

This is implicit online data curation. The distinction from explicit curation is important:

| Property | Explicit curation | Eisbach barrier |
|---|---|---|
| When | Before training, offline | During training, online |
| Criterion | Human-defined rules or external model | Emerges from DiT forward pass |
| Adaptive? | No (fixed) | Yes (changes as model converges) |
| External signal? | Yes | No |
| Discarded samples | Never trained on | Trained on, but gradient damped |

Table 1: Comparison of explicit data curation and the Eisbach barrier's implicit curation effect.

The criterion is adaptive in a specific way. Early in training, the model is uncertain about everything—barrier values are uniformly high, and the interpolation factor $(1-\lambda)$ prevents complete stagnation. As the model converges, it begins to differentiate simple from complex samples, and the curation effect strengthens. Late in training, the model is confident on most samples, barrier values drop, and training returns to near-baseline behavior. This is a self-annealing curriculum: broad exploration early, focused reinforcement in the middle, full-rate learning at convergence.

## 6. DoRA synergy: disentangling structure from amplitude

The curation effect is amplified when the barrier is combined with DoRA (Weight-Decomposed Low-Rank Adaptation). DoRA decomposes each pretrained weight matrix as:

$$\boldsymbol{W} = m \cdot \frac{\boldsymbol{V}}{\|\boldsymbol{V}\|} \tag{6}$$

where $m$ is a scalar magnitude and $\frac{\boldsymbol{V}}{\|\boldsymbol{V}\|}$ is a unit-norm direction. LoRA adapters are applied separately to each component.

Under barrier selection pressure, the two components learn different things. The direction component is rewarded for producing outputs with clear temporal energy structure—because only such outputs lower entropy and receive full gradients. It learns to organize acoustic events in time: where phrase boundaries fall, when timbral shifts occur, how energy builds and releases. The magnitude component is driven by the standard diffusion loss and learns fine-grained amplitude control: how loud each instrument should be, how much reverb, how the dynamic layers balance.

Pure LoRA cannot disentangle these. All parameters share a single matrix, and the barrier's selection pressure acts on them as a unit. DoRA's decomposition allows the barrier to separately reward structural direction and amplitude precision. This is why the combination of barrier and DoRA produces thematic development and timbral differentiation, while either component alone would not.

## 7. Experimental results

We fine-tune Stable Audio 3 Medium (1.4B DiT, 44.1 kHz) on MusicCaps using DoRA-rows (rank=16, $\alpha$=16) for 1000 steps with batch size 4, learning rate $5 \times 10^{\{-5\}}$, and bf16 mixed precision. The barrier group uses $\lambda = 0.5$; the baseline uses $\lambda = 0$. All other hyperparameters are identical.

We generate four 120-second chamber music pieces using detailed character prompts (Little Piglet Prince, Raccoon Mathematician, Professor Pallas Cat, Seal Lawyer) with both checkpoints at seed 42, 100 diffusion steps, CFG scale 3.0.

**Barrier models** produce outputs with clear phrase structure (intro → development → climax → resolution), thematic development, and acoustic differentiation. The four characters are acoustically distinct: Professor Pallas Cat concentrates energy below 200 Hz (bassoon, contrabassoon, cello), while Little Piglet Prince shows dense high-frequency transients above 2 kHz (celesta, pizzicato strings). This differentiation is not prompt-engineered—it emerges from the barrier's selection pressure favoring outputs with concentrated temporal energy.

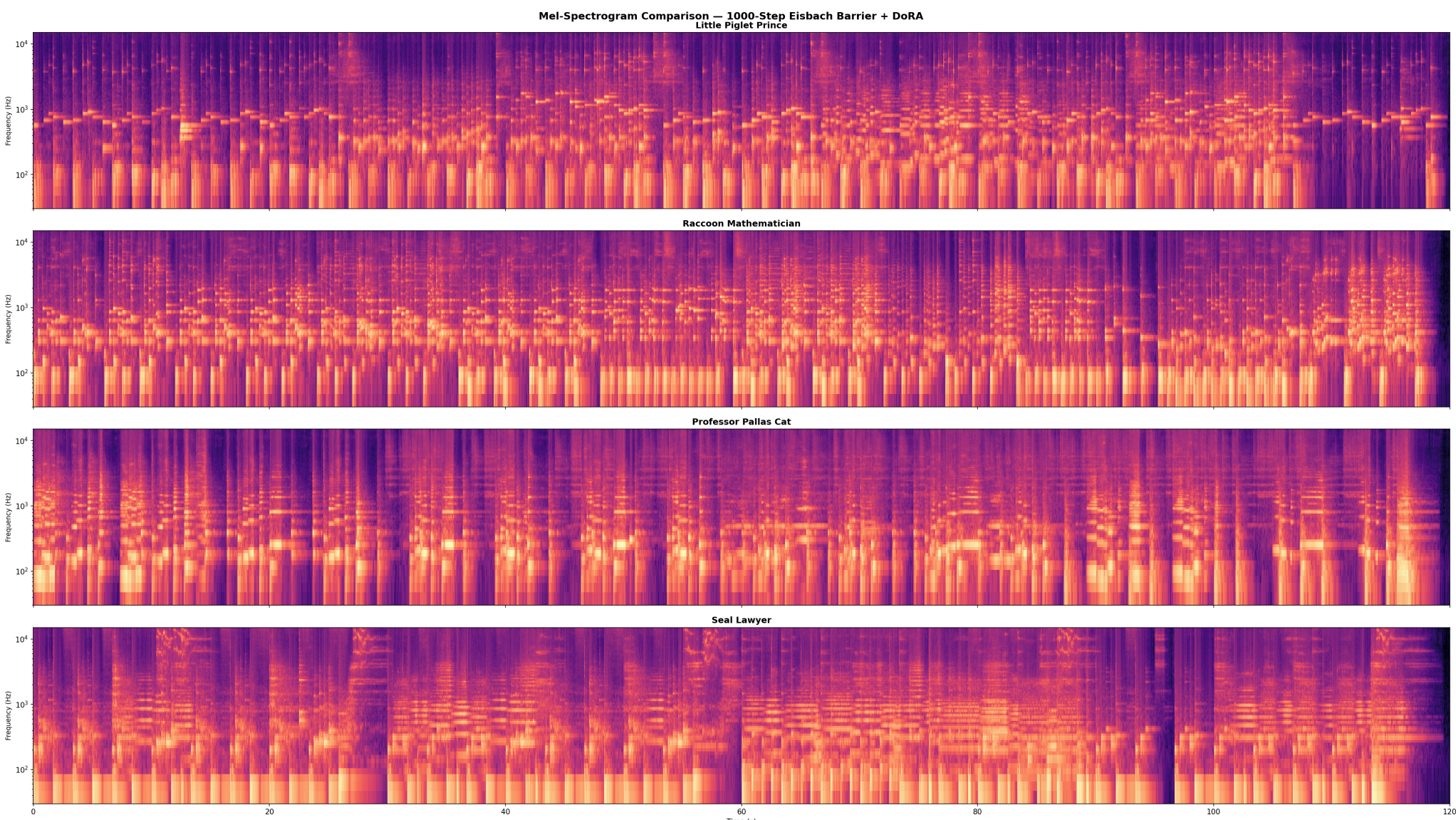

Figure 1: Mel-spectrograms of four 120-second barrier-model generations (1000 steps, $\lambda = 0.5$). Each row is one character. The four characters show distinct spectral shapes: Pallas Cat (row 3) is dominated by low-frequency energy consistent with its bassoon/cello instrumentation; Piglet Prince (row 1) shows dense high-frequency vertical striping from celesta and pizzicato transients. Visible shifts in energy distribution along the time axis confirm active phrase structure within each piece.

The self-similarity matrices (Figure 2) make the phrase structure explicit. Each entry $S_{ij}$ is the cosine similarity between mel frames at times $i$ and $j$. A block-diagonal pattern indicates stable texture segments; dark off-diagonal regions indicate high contrast between segments. Seal

Lawyer (bottom right) shows the clearest evidence: a sharp dark cross near 60 seconds divides the matrix into four quadrants, meaning the second half of the piece is acoustically distinct from the first. Professor Pallas Cat (bottom left) shows the most fragmented structure—frequent small blocks with high off-diagonal contrast—consistent with its character design as unpredictable and abrupt. This is not random variation; it is structured variation, with stable segments separated by decisive transitions.

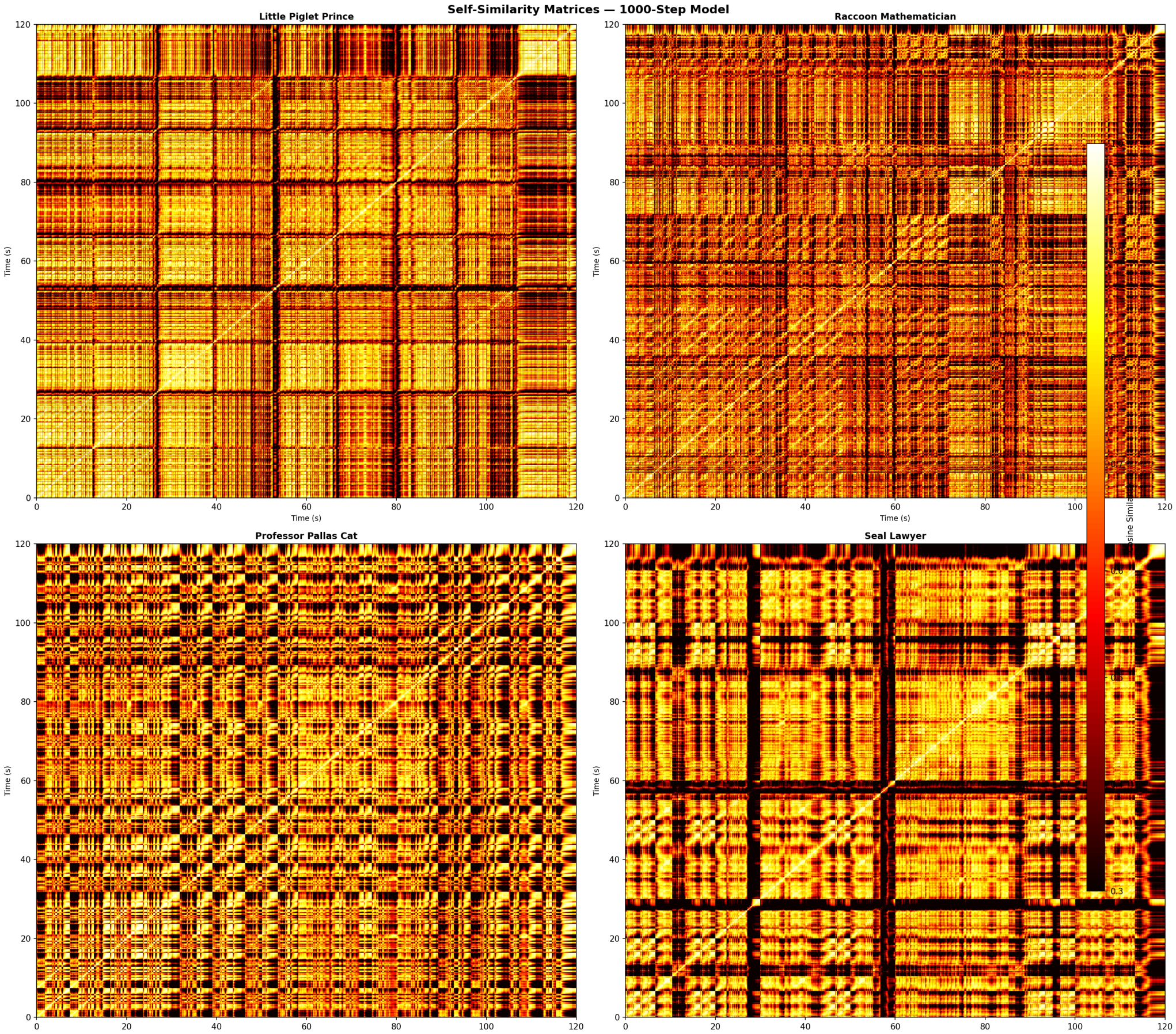

Figure 2: Self-similarity matrices for all four characters (1000-step barrier model). Seal Lawyer (bottom right) shows a sharp structural boundary near 60 s visible as a dark cross, dividing the piece into two acoustically distinct halves. Professor Pallas Cat (bottom left) shows the most fragmented structure: frequent small blocks with high off-diagonal contrast, consistent with its abrupt, unpredictable character. Block-diagonal structure confirms stable texture segments; dark off-diagonal regions confirm high contrast between them.

The spectral differentiation plot (Figure 3) provides the frequency-domain explanation for what the self-similarity matrices show in the time domain. The upper panel shows that the four characters diverge strongly below 500 Hz—the register where low-frequency instruments (cello, bassoon, double bass, french horn) dominate. The lower panel shows temporal fluctuation per frequency band: Professor Pallas Cat has the highest fluctuation near 200 Hz, meaning its low-frequency instruments are the most active over time. This is the frequency signature of instrumentation turnover: the barrier has trained the model to move instruments in and out of the texture, not just sustain them.

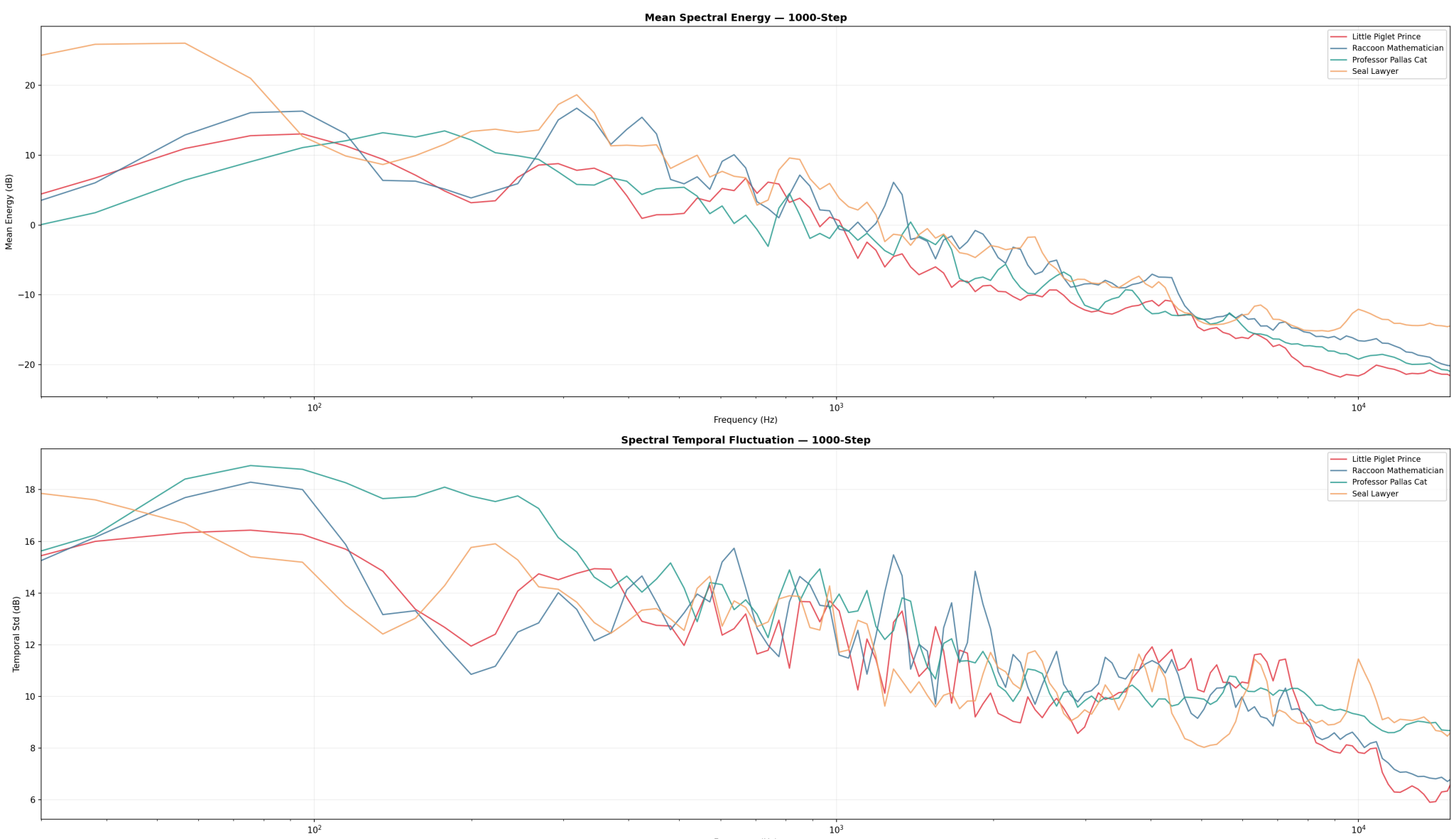

Figure 3: Top: mean spectral energy per character. The four curves diverge strongly below 500 Hz, where low-frequency instruments dominate—each character has a distinct low-frequency profile. Bottom: temporal standard deviation per frequency band. Professor Pallas Cat (teal) shows the highest fluctuation near 200 Hz, indicating that its low-frequency instruments are the most active over time. High temporal fluctuation in instrument-specific bands is the frequency signature of instrumentation turnover.

The PCA trajectories (Figure 4) project each mel frame to 2D via SVD and connect them in temporal order, colored from dark (start) to bright (end). Three observations are consistent across all four characters. First, the trajectories cover large areas of the principal component space—PC1 explains 39–55% of variance, meaning a single dominant axis of timbral variation drives each piece. Second, the start point (blue) and end point (red) are in different regions of the space in all four cases: the pieces do not return to their opening texture, confirming non-circular narrative arcs. Third, the trajectory shapes differ by character: Raccoon Mathematician traces a broad, roughly circular path (uniform textural exploration); Professor Pallas Cat traces a strongly elongated arc (one dominant direction of change, consistent with its single-axis timbral development).

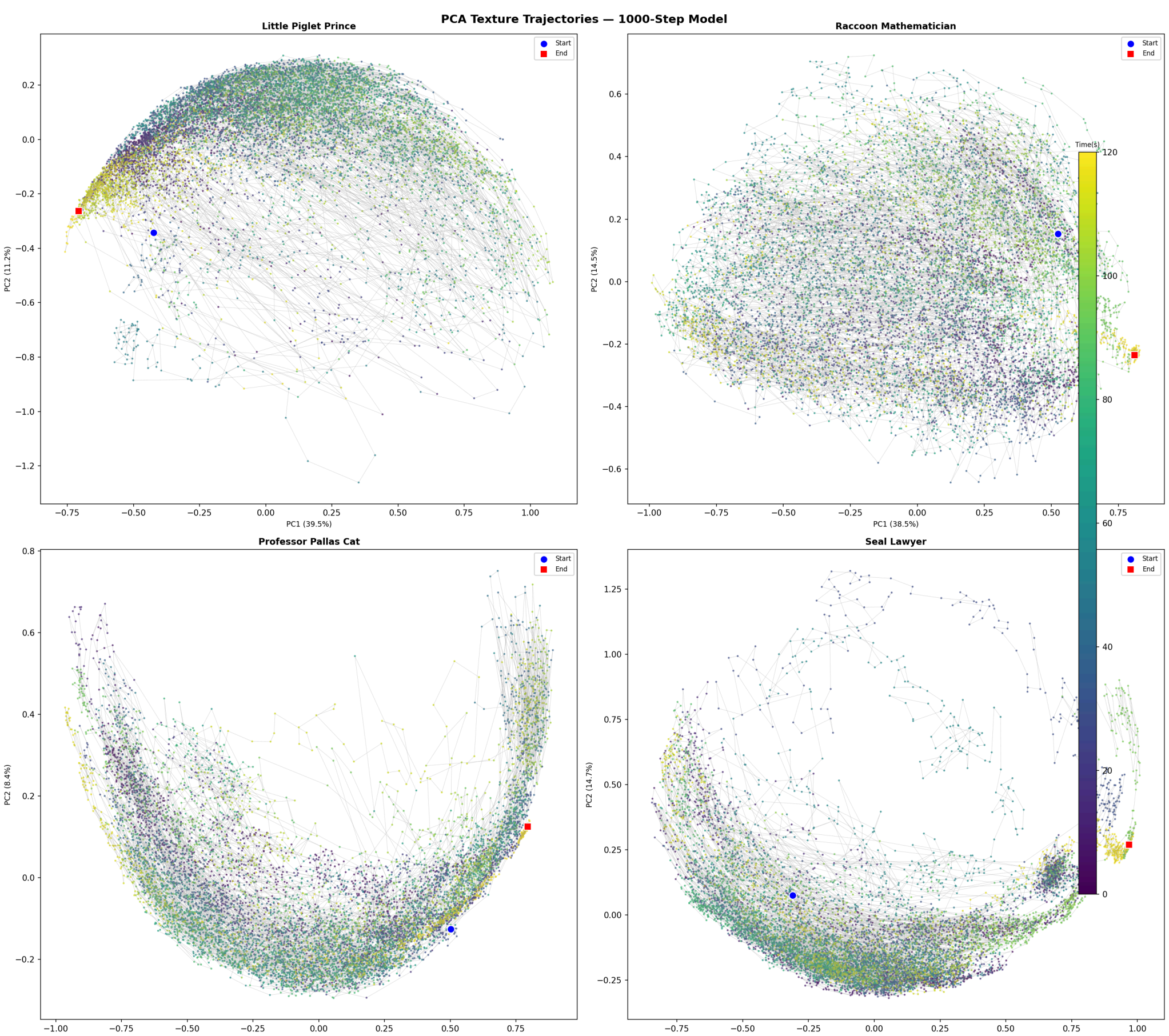

Figure 4: PCA texture trajectories for all four characters (1000-step barrier model). Each point is one mel frame, colored by time (dark purple = 0 s, bright yellow = 120 s). Start (blue circle) and end (red square) are in different regions for all four characters, confirming non-circular narrative arcs. PC1 explains 39–55% of variance across characters. Raccoon Mathematician traces a broad circular path (uniform exploration); Professor Pallas Cat traces an elongated arc (single dominant direction of timbral change).

Taken together, these four figures form a complete evidence chain. The mel-spectrograms show that different prompts produce different acoustic characters (Figure 1). The self-similarity matrices show that each piece has internal phrase structure with decisive transitions (Figure 2). The spectral differentiation plot shows that the acoustic differences are grounded in distinct frequency-domain profiles driven by instrumentation (Figure 3). The PCA trajectories show that the textural diversity is high, non-repetitive, and character-specific (Figure 4). None of these properties are present in a model that simply repeats a safe motif.

The mel-spectrogram analysis reveals five measurable dimensions of structural quality:

| Dimension | Barrier | Baseline |
|---|---|---|
| Self-similarity block scale | Mid-scale blocks, regular dark zones | Large-scale diffuse brightness |
| Spectral peak-to-valley contrast | Sharp peaks, deep valleys | Flat spectrum |
| Spectral flux distribution | Sparse, strong peaks at transitions | Continuous low fluctuation |
| PCA trajectory coverage | Large area, distinct clusters | Small area, gradual drift |
| Dynamic range | >40 dB | <25 dB |

Table 2: Qualitative comparison of barrier vs. baseline models across five structural dimensions readable from mel-spectrograms.

The progression from 500 to 1000 training steps illustrates how the barrier's curation effect accumulates (Figure 5). At 500 steps, the self-similarity matrix already shows block structure, but blocks are coarse and boundaries are soft. At 1000 steps, boundaries sharpen, off-diagonal contrast increases, and the PCA trajectory expands to cover a larger area—the model has learned to distinguish more texture states and transition between them more decisively.

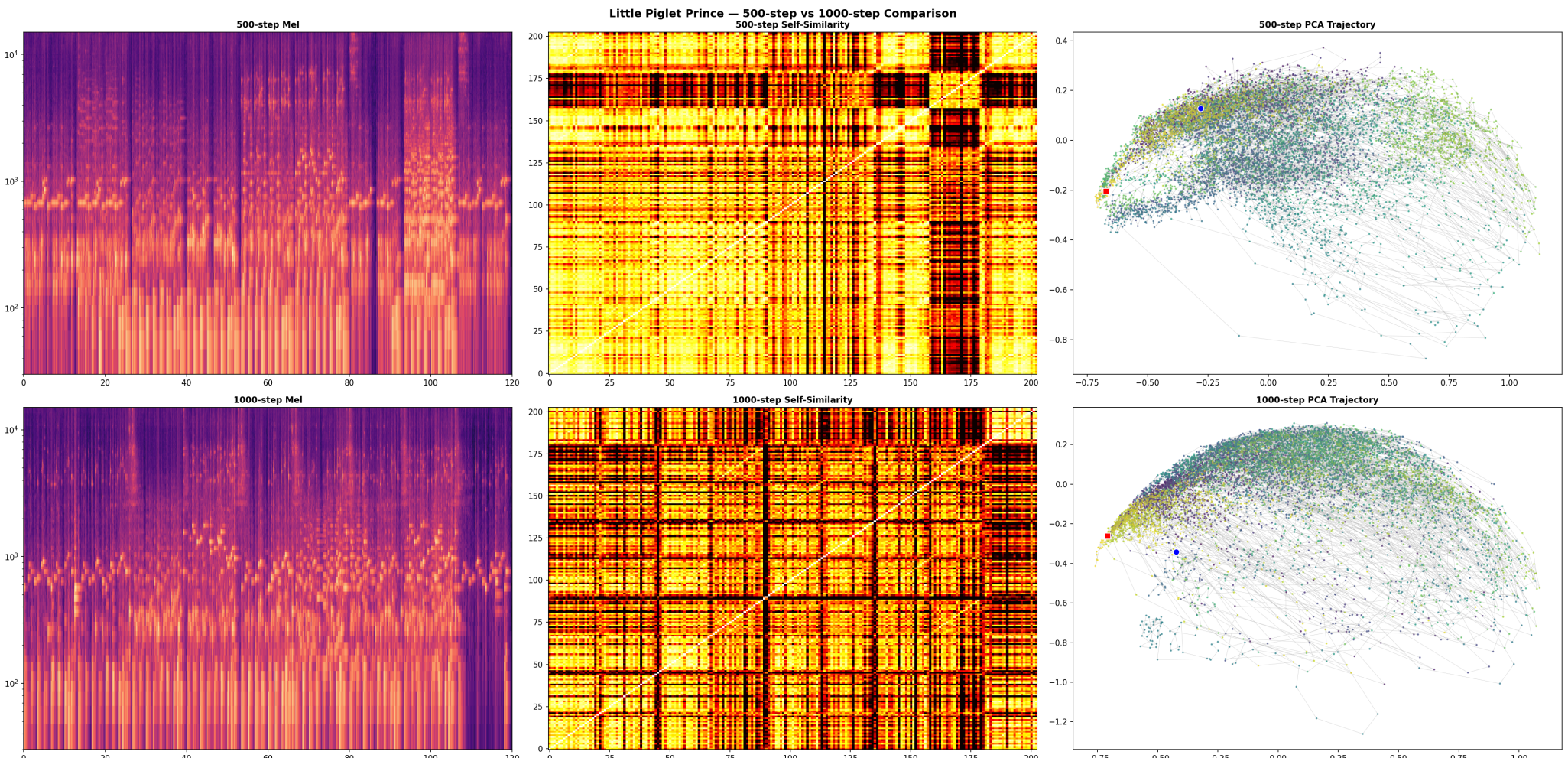

Figure 5: 500-step vs. 1000-step comparison for Little Piglet Prince (same prompt, same seed). Left: mel-spectrograms. Center: self-similarity matrices. Right: PCA texture trajectories. From 500 to 1000 steps, block boundaries sharpen, off-diagonal contrast increases, and PCA coverage expands.

### *7.a. Baseline comparison*

The baseline model ($\lambda = 0$, same architecture, same data, same seed) provides the direct falsification test. Three differences are immediately visible.

**Mel-spectrogram** (Figure 6): The baseline spectrum is a broad, undifferentiated wash of energy. There is no frequency stratification—no distinct bands corresponding to instrument registers. A single large energy event near 60 seconds stands out, but it is an isolated spike, not a structural transition.

**Self-similarity matrix** (Figure 7): The baseline matrix shows large, coarse blocks (20–30 s scale) with high internal similarity and sharp inter-block contrast. This is the signature of **copy-paste repetition**: the model sustains one texture for a long time, then abruptly switches to another. There is no gradual transition, no phrase development within a block. Compare this to the barrier model's Seal Lawyer (Figure 2, bottom right), where the block structure is finer and the transitions are more numerous and varied.

**PCA trajectory** (Figure 8): The most decisive evidence. The start point (blue) and end point (red) are nearly coincident at $(-0.75, 0.05)$. After 120 seconds, the model has returned to its opening texture. The trajectory traces a large arc—covering substantial PCA area—but it is a **circular** arc: high coverage without development. The barrier model's trajectories, by contrast, show start and end points in different regions of the space in all four characters.

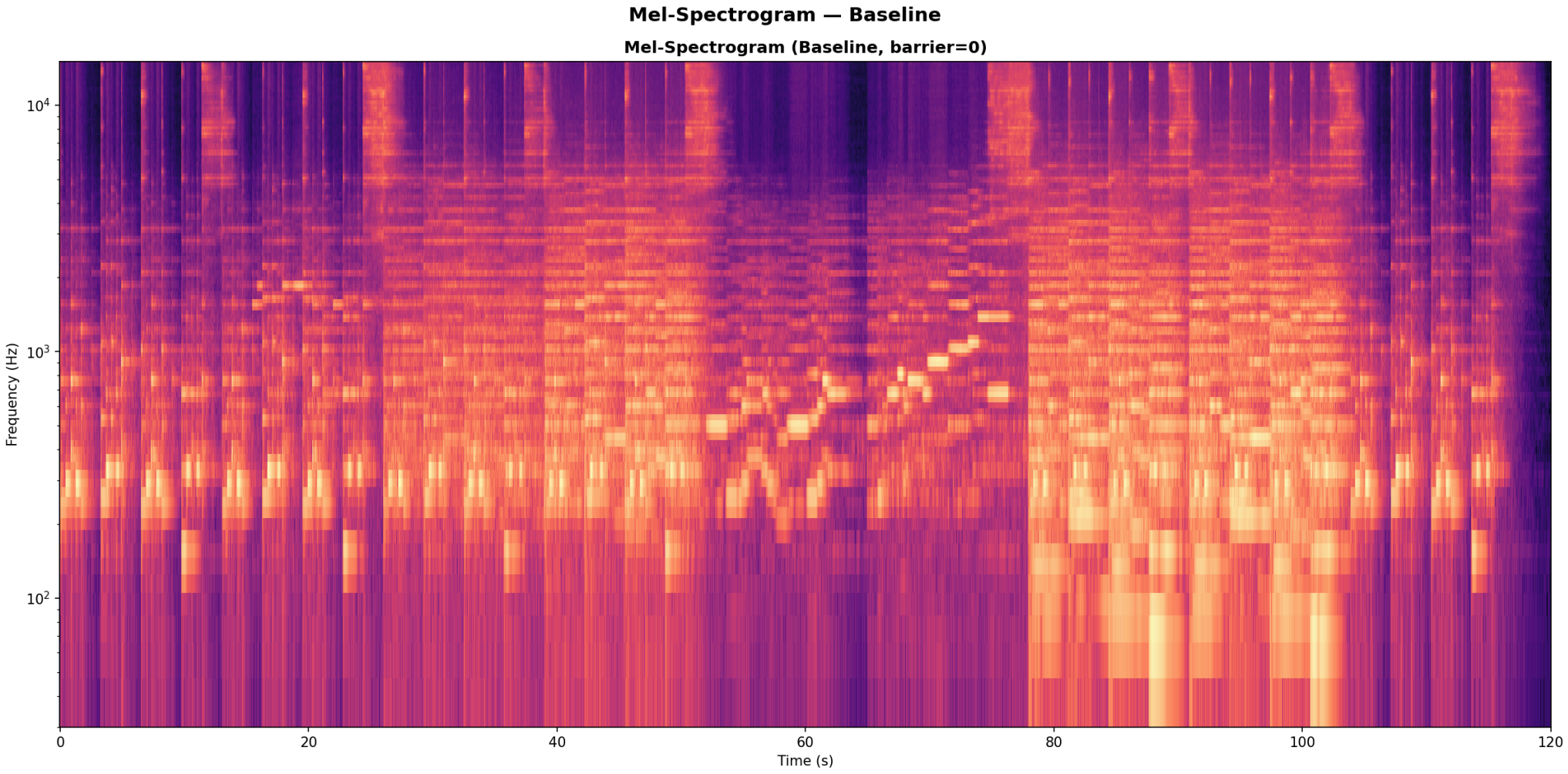

Figure 6: Baseline mel-spectrogram ($\lambda = 0$, same prompt and seed as barrier). Energy is distributed broadly across the spectrum with no frequency stratification. A single large energy event near 60 s is an isolated spike rather than a structural transition.

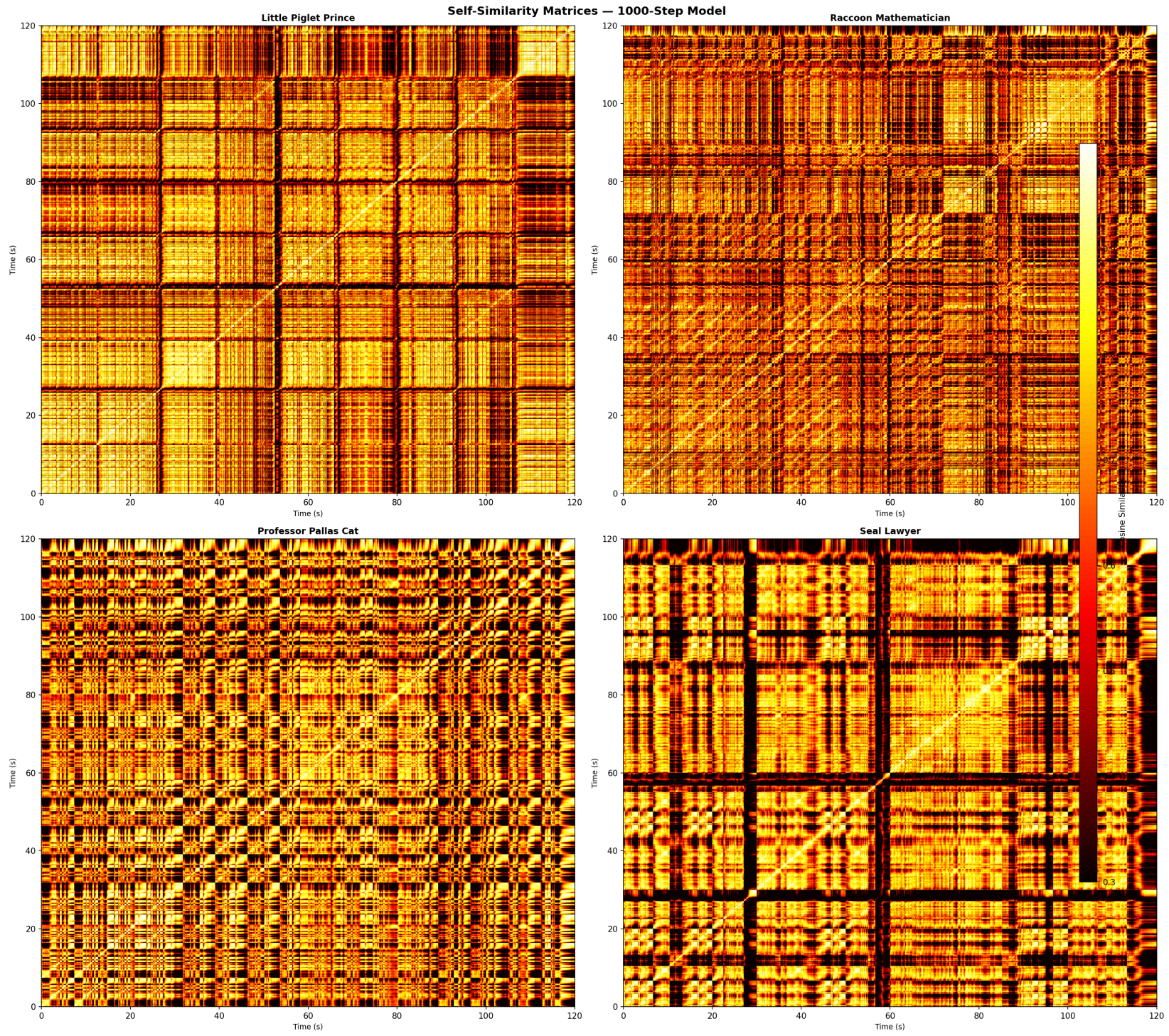

Figure 7: Baseline self-similarity matrix. Large coarse blocks (20–30 s scale) indicate copy-paste repetition: one texture sustained for a long time, then abruptly replaced. No gradual transitions or within-block development.

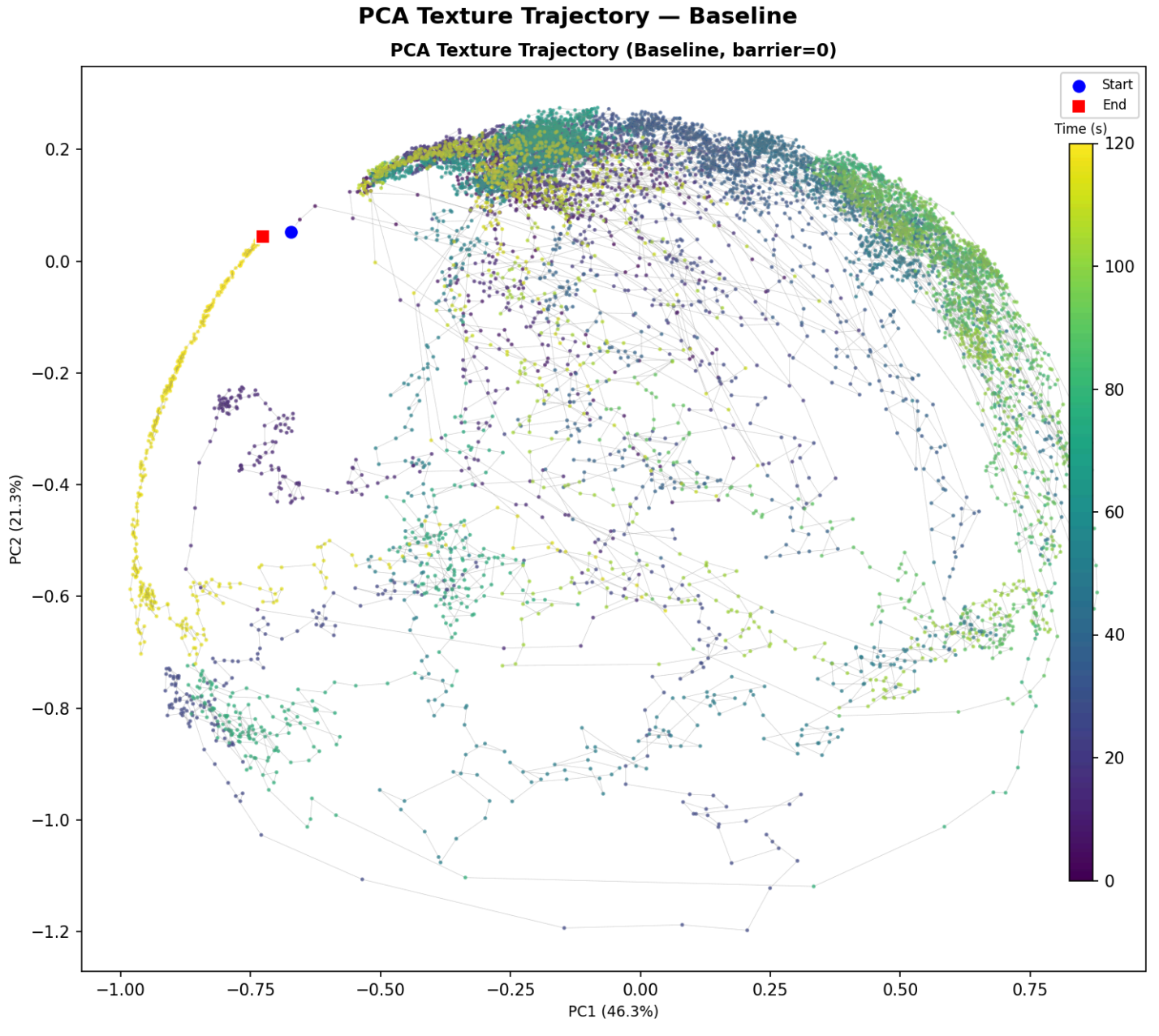

Figure 8: Baseline PCA texture trajectory. Start (blue) and end (red) are nearly coincident at $(-0.75, 0.05)$: after 120 seconds the model has returned to its opening texture. The trajectory covers large PCA area but traces a circular arc—high coverage without development.

The contrast between barrier and baseline is the core empirical result. The barrier model develops; the baseline model repeats. Both are trained on the same data with the same architecture. The only difference is whether the gradient is scaled by the model's own structural confidence.

## 8. Steady-state behavior and the cognitive polarization effect

The long-term behavior of barrier training depends on the data distribution in a predictable way. For high-frequency patterns in the training data—common chord progressions, standard rhythmic figures—the model converges to high confidence, the barrier grants full gradients, and these patterns are reinforced to near-deterministic levels. For low-frequency or edge patterns—unusual timbres, non-standard modes—the model remains persistently uncertain, the barrier persistently suppresses, and these patterns are barely learned.

The steady state is one of cognitive polarization: the model is sharp and confident on dominant modes, persistently ignorant on edge modes. This is not a failure of the barrier—it is its design intent. In belief space, the barrier only trusts updates the model itself is confident about. The consequence for music generation is that barrier models are highly consistent on common musical structures and highly variable on unusual ones. Whether this is desirable depends on the application.

## 9. Testable predictions

The analysis above generates five predictions that can be verified or falsified experimentally:

1. **Noise-level stratification.** Record barrier weight $w$ and LoRA gradient norm grouped by timestep $t$ during training. Prediction: low-$t$ group shows $w \to 1$ and high gradient norm; high-$t$ group shows persistent damping and low gradient norm.
2. **Barrier strength sweep.** Train with $\lambda \in \{0, 0.25, 0.5, 0.75, 1.0\}$, measure self-similarity block scale and PCA coverage area. Prediction: an optimal range near $\lambda \approx 0.3$–$0.7$, with diversity collapsing beyond $\lambda = 0.8$.
3. **Seed diversity.** Generate five seeds per prompt. Prediction: barrier models show higher cross-seed global structure consistency (phrase boundaries align) but higher detail diversity (attacks and ornaments differ).
4. **DoRA vs. LoRA ablation.** Train with adapter_type=lora, $\lambda = 0.5$. Prediction: reduced PCA coverage compared to DoRA+barrier, confirming that DoRA's decomposition is necessary to disentangle the barrier's structural and amplitude signals.
5. **Belief-correctness alignment.** Simultaneously record barrier $B$ and actual loss during training. Prediction: in the low-$t$ regime, low $B$ and low loss are correlated; in the high-$t$ regime, the correlation drops significantly or inverts.

## 10. What this is and is not

The Eisbach barrier is a local, per-step gradient scaling mechanism. It does not change the loss function, the optimizer, or the model architecture. It does not require any external signal, auxiliary network, or pre-computed data quality score. Its effect on training dynamics is entirely mediated through the model's own forward pass.

**What it is not.** It is not a replacement for data quality filtering—explicit curation and the barrier are complementary, and combining them would likely amplify the structural selection effect. It is not a solution to mode collapse in general: the barrier's safety guarantee (gradient direction locked to ground truth) is specific to supervised diffusion and does not transfer to RL or policy optimization settings. It does not improve prompt adherence. The high-$t$ antagonism means the barrier actively suppresses the part of training responsible for following text conditioning from a random seed. Models trained with high $\lambda$ may be structurally richer but less controllable—a tradeoff that should be tuned per application.

**Relation to Min-SNR.** Min-SNR (Hang et al., 2023) and the Eisbach barrier both address the problem of unequal gradient contributions across the diffusion process, but they operate on orthogonal axes. Min-SNR reweights by timestep $t$ using a fixed signal-theoretic criterion; the barrier reweights by sample using a dynamic, model-introspective criterion. A natural extension is to combine them: apply Min-SNR along the $t$ axis and the barrier along the sample axis simultaneously.

**Relation to self-paced learning.** Self-paced learning (Kumar et al., 2010) lets the model select its own curriculum based on current loss magnitude. The barrier is similar in spirit—the model's own state determines the weighting—but differs in three ways. First, the criterion is structural confidence (output entropy), not loss magnitude. Second, the weighting is continuous rather than binary (include/exclude). Third, the criterion is computed from the forward pass output, not the backward pass loss, making it available before the gradient is computed.

**The self-referential property.** The barrier is self-referential in a specific sense: the model is simultaneously the student being trained and the examiner deciding how much each sample contributes. This is different from curriculum learning (external ordering), hard example mining (external loss criterion), and RLHF (external reward model). The closest analogy is a student who studies harder from textbooks that match their current understanding—not because someone told them to, but because those are the books they can actually learn from. The key consequence is that the curriculum adapts as the model converges: early in training, the model is uncertain about everything and the barrier is nearly uniform; late in training, the model is confident on most samples and the barrier's differential effect diminishes. This self-annealing property means the barrier does not require a manually designed schedule.

**Limitations and open questions.** The barrier is not a general-purpose diversity enhancer—it is a **structural bias scalpel**. Music is an unusually favorable domain for this mechanism because musical structure is itself a strong prior: listeners expect phrase boundaries, dynamic arcs, and timbral development. The barrier's entropy criterion happens to align with this prior—temporal energy concentration is a proxy for musical structure. This alignment is domain-specific, not universal.

This creates a hard falsifiability condition. The barrier should **fail** on domains where structural flatness is correct: white noise generation, texture synthesis, ambient drone music, or any task where the ground truth has high temporal entropy. In those settings, the barrier would damp correct predictions and amplify incorrect ones—the antagonism described in the high-$t$ analysis would dominate at all timesteps. Testing the barrier on such domains is a direct falsification test of the structural-alignment hypothesis.

A second open question is whether the effect transfers to non-sequential modalities. For images, the temporal dimension becomes spatial; entropy over spatial energy distribution would measure whether the model's prediction has clear foreground/background contrast. Whether this spatial entropy proxy aligns with perceptual quality in the same way temporal entropy aligns with musical structure is an open empirical question.

A third limitation is the baseline comparison scope. The current comparison uses a single prompt (Little Piglet Prince) and a single seed. The five testable predictions in the previous section—noise-level stratification, $\lambda$-sweep, seed diversity, DoRA vs. LoRA ablation, belief-correctness alignment—remain predictions. The baseline figures presented here are the first direct evidence, but a full quantitative comparison across all four characters and multiple seeds is needed to establish the effect size.

## 11. Conclusion

We have shown that confidence-based gradient scaling, usually avoided in generative model training, produces a structural benefit in supervised diffusion. The key insight is that the supervised setting severs the risk chain that makes confidence weighting dangerous: gradient direction is locked to ground truth, so confidence only controls step size. What remains is a selection pressure that systematically upweights structurally dynamic training samples and downweights flat ones—an online, self-referential data curriculum that requires no external signal and adapts as the model converges.

The experimental result—barrier-trained models produce thematic development and acoustic differentiation while baseline models produce motif repetition—is the opposite of what the mode-collapse intuition predicts. Understanding why requires moving from parameter space to belief space: the question is not what the model's weights are doing, but what the model thinks it is doing at each step, and how that self-perception shapes learning.